\documentclass [a4paper,12pt]{article}
\usepackage{amssymb}
\usepackage{graphicx}
\usepackage{amsmath}
\usepackage{fancybox}
\usepackage{color}
\usepackage{geometry}
\usepackage{mathtools}
\usepackage{empheq}

\newcommand{\be}{\begin{equation}}
\newcommand{\ee}{\end{equation}}
\newcommand{\CS}{\text{CS}} 
\newcommand{\SP}{\text{sp}} 
\newcommand{\EQREF}[1]{eq.\eqref{#1}} 
\begin{document}
\setlength{\baselineskip}{18pt}
\begin{titlepage}

\begin{flushright}
NITEP 149
%OCU-PHYS 574
%KOBE-TH-19-06       
\end{flushright}
\vspace{1.0cm}
\begin{center}
{\LARGE\bf Analytic construction of sphaleron-like solution invoking higher dimensional gauge theory}  
\end{center}
\vspace{15mm}

\begin{center}
{\large

Yuki Adachi, C.S. Lim$^a$ and Nobuhito Maru$^{b, c}$
}
\end{center}

\vspace{1cm}

\centerline{{\it
Department of Sciences, Matsue College of Technology,
Matsue 690-8518, Japan.}}

\centerline{{\it
$^{a}$
Tokyo Woman's Christian University, Tokyo 167-8585, Japan }} 

\centerline{{\it 
$^b$
Department of Physics, Osaka Metropolitan University,
Osaka 558-8585, Japan}}

\centerline{{\it 
$^c$
Nambu Yoichiro Institute of Theoretical and Experimental Physics (NITEP), }} 
\centerline{{\it Osaka Metropolitan University,  
Osaka 558-8585, Japan}}
%
%
%   Abstract

\vspace{1.5cm}
%\centerline{\large\bf Abstract}
\vspace{0.5cm}
 \begin{abstract}

We perform analytic construction of a sphaleron-like solution in the 4-dimensional 
(4D) space-time invoking  the framework of 5D SU(2) gauge theory. 
By the sphaleron-like solution we mean a static finite energy solution to the equation of motion, which carries the Chern-Simons number $N_\CS  = \frac{1}{2}$. 
Since we are interested in the static solution in the low-energy effective theory, 
we focus on the part of the action which contains only the gauge fields in the 4D space (not space-time),
$(A_{i}, A_{y}) \ (i = 1,2,3)$ and keep only the Kaluza-Klein zero modes of these fields. 
Interestingly, the self-duality condition in this 4D space is known to be nothing but the BPS condition for the 't Hooft-Polyakov monopole,
once the extra-space component $A_y$ is identified with the adjoint scalar, needed for the monopole solution. 
Thus, the sphaleron-like solution is based on the BPS monopole embedded in the higher dimensional space-time,
which may be interpreted as a self-dual gauge field. 
By use of the lesson we learn in the case of the instanton in ordinary 4D space-time, 
we achieve the sphaleron-like configuration of $A_{i}$, which carries $N_\CS  = \frac{1}{2}$.
As a characteristic feature of this construction invoking higher dimensional gauge theory, in clear contrast to the case of the ordinary BPS monopole, 
the VEV of the adjoint scalar is topologically fixed, and therefore the mass of the sphaleron-like solution is determined to be $M_\SP  = \frac{4\pi}{g_{4}^{2}}\frac{1}{R}$ ($g_{4}$: 4D gauge coupling constant, 
$R$: the radius of the circle as the extra space).
We also argue that the sphaleron-like solution may be regarded as a saddle point of the energy in the space of static field configurations. 

 \end{abstract}

\end{titlepage}

\section{Introduction}

One of the most important problems to be addressed in the cosmology is how the matter, existing in the present universe, 
is generated starting from the early universe where the same amount of matter and anti-matter was present.
The standard model (SM) of elementary particle meets the Sakharov's necessary conditions for \lq\lq baryogenesis\rq\rq .
Namely, the weak interaction of quarks breaks CP symmetry.
Also, although the action itself preserves baryon number B, B conservation is violated by 't Hooft anomaly \cite{'t Hooft}.

Unfortunately, the probability of the B violating process described by the instanton configuration of the gauge field, connecting vacua having different topological winding number, i.e. the Chern-Simons (CS) number $N_\CS $, is highly suppressed by a factor $e^{- \frac{16\pi^{2}}{g^{2}}}$.
However, under the circumstance of finite temperature $T$, it is replaced by a \lq\lq Boltzmann\rq\rq factor $e^{-\frac{M_\SP }{T}}$ and we expect enough transition probability of the B violating process for higher temperatures.
Here, $M_\SP $ stands for \lq\lq sphaleron mass\rq\rq, the energy possessed by the field configuration of the SU(2) gauge field and the Higgs field, called sphaleron, which is regarded as the field configuration corresponding to the 
intermediate state of the process connecting neighboring vacua with CS numbers different by one unit. Thus, the sphaleron is characterized as the field configuration, which carries $N_\CS  = \frac{1}{2}$. 

The sphaleron configuration was found as a static solution to the equation of motion in the system of SU(2) gauge field and Higgs 
doublet, in the limiting case of vanishing weak mixing, $\theta_{W} = 0$, of the SM \cite{Manton}. 
Though it has been argued that the SM does not provide enough amount of baryon number in the universe, the sphaleron  plays a crucial role also in the scenario of \lq\lq leptogenesis\rq\rq, where it has a function to convert the lepton number $L$ generated by the Majorana masses of neutrinos in higher energy scale into the baryon number $B$ \cite{Fukugita Yanagida}. 

In the construction of the sphaleron solution, however, the solution was found numerically, since the coupled field equations for the gauge and Higgs fields are not easy to solve analytically, while the instanton solution is analytically constructed as the solution to the self-duality condition.
(For a  numerical construction of the sphaleron solution based on a well-devised method by use of gradient flow, see \cite{Hamada}.) 
Although the numerical result is sufficient for the purpose to estimate $M_\SP $, it will be nice if we can construct the solution analytically, in order to get deeper insight into the property of this important field configuration.

From such a point of view, in this paper, as the first attempt towards the analytic construction of the sphaleron solution, we propose analytic construction of sphaleron-like solution in the coupled system of SU(2) gauge field 
and Higgs field which belongs to the adjoint (triplet) representation of SU(2).
More precisely, we construct analytically a static solution with finite energy to the equation of motion in the system, with the gauge field carrying $N_\CS  = \frac{1}{2}$. 

As we will see below, in the construction a key ingredient is invoking the framework of higher dimensional gauge theory, especially the scenario of gauge-Higgs unification (GHU).
GHU is a scenario of the physics beyond the standard model (BSM), where 
the Higgs field is identified as the extra-space component of the higher dimensional gauge field \cite{Manton:1979kb}, \cite{Hosotani}. 
In the scenario the well-known gauge hierarchy problem is naturally solved by virtue of the higher dimensional gauge symmetry \cite{Hatanaka:1998yp}.
Also, since the Higgs field may be regarded as a Wilson-line phase, in the S$^{1}$ compactification of five-dimensional (5D) GHU, for instance, the Higgs potential is periodic in the Higgs field, which leads to characteristic implications, not shared by other types of BSM models, for Yukawa couplings \cite{HKLT}, the strong CP problem \cite{ALM1} and the structure of the vacuum state \cite{ALM2}, etc.     

Since the instanton in ordinary 4D space-time is the gauge field configuration which describes the transition between the vacuum with $N_\CS  = n \ (n: {\rm integer})$ at $t = -\infty$ and the neighboring vacuum with $N_\CS  = n+1$ at $t = \infty$, it will be a natural guess that its time slice at $t = 0$ is a good candidate for the analytic construction of the gauge field with $N_\CS  = \frac{1}{2}$.
Unfortunately, this possibility faces a serious problem: the energy of the time slice is dependent on a free parameter of the solution, i.e. the size of the instanton, so the $M_\SP $, and therefore the transition probability between these vacua at 
finite temperature is not predictable \cite{Manton_textbook}. 

This problem is found to be solved in the framework of higher dimensional gauge theory.
In this paper we work in the 5D SU(2) GHU model.
Since we are interested in the static solutions, we focus on the 4D space (not ordinary space-time) with coordinates $(\vec{x}, y)$, with $y$ being the extra space coordinate, ignoring the time coordinate $t$.
We also concentrate on the Kaluza-Klein (KK) zero modes of the fields, which are expected to describe the low-energy effective theory.
Then, interestingly the (anti-)self-duality condition for the gauge fields $(\vec{A}, A_{y})$ in the 4D space is found to be nothing but the BPS condition for the 't Hooft-Polyakov monopole (TPM) \cite{'t Hooft_monopole, Polyakov}, once the extra space component of the gauge field $A_{y}$ is identified with the adjoint Higgs field, necessitated for the construction of the monopole solution.

Thus, by learning the way to construct the gauge field configuration with $N_\CS  = \frac{1}{2}$ in the case of instanton, we succeed in constructing a static finite energy solution to the equation of motion with the gauge field carrying $N_\CS  = \frac{1}{2}$, \lq\lq sphaleron-like\rq\rq solution.
Namely, the sphaleron-like solution is the TPM embedded in the framework of GHU.
To be more concrete, we first perform suitable $y$-dependent local gauge transformation for the monopole solution in order to eliminate $A_{y}$, and then take the slice at $y= 0$ of $\vec{A}$ to realize $N_\CS  = \frac{1}{2}$.

One remarkable feature of this construction is that the vacuum expectation value (VEV) of the adjoint scalar $A_{y}$ is not a free parameter but is quantized, just as the action of the self-dual gauge field is topologically quantized by the winding number, though in the case of the ordinary BPS monopole the VEV is left as a free parameter. Thus the energy possessed by the field configuration, namely the mass of the sphaleron-like solution, $M_\SP $, is topologically fixed and now the transition probability between the neighboring vacua at finite temperature becomes predictable, in clear contrast to the case of ordinary instanton.   

We also discuss the instability of the sphaleron-like solution: we argue that it may be regarded as a saddle point of the energy in the space of static field configurations, where the energy decreases along the direction in which $N_\CS $ changes, just as in the case of ordinary instanton, by showing two concrete examples of the parameter-dependent field deformation.

The BPS monopole may also be understood as a self-dual field configuration, once the time direction is compactified on S$^{1}$ in the 4D space-time and if we ignore the $t$-dependence of the gauge fields, or 
equivalently if we consider the high temperature limit of the finite temperature SU(2) gauge theory.  
In fact, in the literature there exist some discussions on the solution to the self-dual condition of the SU(2) gauge field on the R$^{3} \times$ S$^{1}$ Euclidean space, called \lq\lq SU(2) caloron\rq\rq or \lq\lq periodic instanton\rq\rq  \cite{Lee and Lu}, \cite{Kraan and 
van Baal}. In these papers, however, the attempt to construct a gauge field configuration with $N_\CS  = \frac{1}{2}$ does not seem to be made, though it is the main issue in the present paper,

\section{The time slice of the instanton solution}  

In this section, we first learn how to construct a gauge field configuration with CS number $N_\CS  = \frac{1}{2}$ by considering the instanton solution in the ordinary 4D space-time. 
It is well-known that the (one) instanton solution of SU(2) gauge theory formulated in the ordinary 4D space-time,    
\be 
\label{1.1} 
A_{0} = -i \frac{\vec{x}\cdot \vec{\sigma}}{\rho^{2} + \lambda^{2}},~
\vec{A} = i \frac{x_{0}\vec{\sigma}+ \vec{\sigma}\times \vec{x}}{\rho^{2} + \lambda^{2}} ~~ \left(\rho^{2} = x_{0}^{2}+r^{2}, \ r = |\vec{x}|\right),  
\ee 
describes the transition from $N_\CS  = n$ at $t = -\infty$ into $N_\CS  = n+1$ at $t = \infty$ ($n$: integer). 
In \EQREF{1.1}, gauge fields should be understood to be anti-hermitian with the factor $-ig$ ($g$: gauge coupling constant) being included in the gauge fields, and $\lambda$, the size of the instanton, is a free parameter.
We therefore naively expect that the time slice of $\vec{A}$ at $t = 0$ carries $N_\CS  = n + \frac{1}{2}$.
Thus for $n = 0$, it is anticipated that a configuration with $N_\CS  = \frac{1}{2}$ is realized analytically.

In general, CS number $N_\CS $ is given in terms of the gauge field $A_{i}$ and its field strength $F_{ij}$ as, 
\be 
\label{1.2} 
N_\CS  = -\frac{1}{16\pi^{2}} \int d^{3}x \ \epsilon _{ijk} \ {\rm Tr} \left[ A_{i}F_{jk} - \frac{2}{3}A_{i}A_{j}A_{k} \right].
\ee 
To start with, we move to a gauge, where $A_{0}$ vanishes, by a suitable gauge transformation.
Note that in this gauge, among the surface integral of the Chern-Simons current 
$J^{\mu}_\CS \equiv - \frac{2}{g^{2}} \epsilon^{\mu \nu \rho \sigma} {\rm Tr}(A_{\nu}F_{\rho \sigma} - \frac{2}{3}A_{\nu}A_{\rho}A_{\sigma})$ on the cylindrical surface, the integral over the side surface $r = \infty \ (-\infty < t < \infty)$ is vanishing, and only the difference of the surface integrals on the two end cap surfaces, $t = -\infty$ and $t = \infty$, contributes to the winding number 1, which is carried by the instanton itself.

Consider a local gauge transformation by an unitary matrix $V$, 
\be 
\label{1.3} 
A_{\mu} \ \to \ A'_{\mu} = V^{\dagger} A_{\mu}V + V^{\dagger}\partial_{\mu}V. 
\ee 
The gauge $A'_{0} = 0$ is realized by choosing $V$ so that it satisfies 
\be 
\label{1.4} 
\partial_{0}V = - A_{0}V = i \frac{\vec{x}\cdot \vec{\sigma}}{\rho^{2} + \lambda^{2}}V, 
\ee
whose solution is easily found to be 
\be 
\label{1.5}
V = e^{i \theta \vec{\hat{x}}\cdot \vec{\sigma}} \ \ 
\left(\vec{\hat{x}} = \frac{\vec{x}}{r}\right),  
\ee 
where 
\be  
\label{1.6} 
\theta =
\frac{r}{\sqrt{r^{2}+\lambda^{2}}}
\left[ \tan^{-1}\frac{x_{0}}{\sqrt{r^{2}+\lambda^{2}}} + \left(n + \frac{1}{2}\right)\pi \right] .
\ee          
The constant of integral has been chosen to be $(n + \frac{1}{2})\pi$, so that $N_\CS  = n$ at $t = -\infty$. For $t = \pm \infty$, 
$\vec{A}$ vanishes and $A'_{i} =  V^{\dagger}\partial_{i}V$ is easily found to have $N_\CS  = n, \ n+1$, since for pure gauge configurations $F_{jk} = 0$ and 
in general, 
\be 
\label{1.7} 
N_\CS  = \frac{1}{24\pi^{2}} \int d^{3}x \ \epsilon _{ijk} \ {\rm Tr} (A_{i}A_{j}A_{k}) = n \ \ 
\text{for} \ 
A_{i} = U^{\dagger}\partial_{i}U \ \ 
\left(U = e^{i n\pi \frac{r}{\sqrt{r^{2}+\lambda^{2}}}\vec{\hat{x}}\cdot \vec{\sigma}}\right).  
\ee 
Let us note that the vacua with different integers of $N_\CS $ are all equivalent, and the theory has translational invariance, i.e. the invariance under the transformation which changes $N_\CS $ by an arbitrary integer caused by \lq\lq large\rq\rq gauge transformation.
What has a gauge invariant meaning is the difference of $N_\CS $ between distinct vacua, which is nothing but the winding number carried by the instanton solution.

For general $x_{0}=t$, substituting \EQREF{1.1}, \EQREF{1.5} and \EQREF{1.6} in the relation $A'_{i} = V^{\dagger} A_{i}V + V^{\dagger}\partial_{i}V$ we obtain after a little lengthy calculation,  
\begin{equation}
\label{1.8} 
 \begin{split}
  A'_{i} =& i \Bigg[ 
  \left(\frac{\sin 2\theta \ x_{0} + \cos 2\theta \ r}{\rho^{2} + \lambda^{2}} + \frac{1-\cos 2\theta}{2r}\right) e^{1}_{i} 
  + \left(\frac{\cos 2\theta \ x_{0} - \sin 2\theta \ r}{\rho^{2} + \lambda^{2}} + \frac{\sin 2\theta}{2r}\right) e^{2}_{i} \\ 
  &+  \lambda^{2}\left\{\frac{1}{\rho^{2}+\lambda^{2}}\frac{x_{0}}{r^{2}+\lambda^{2}} + \frac{\theta}{r(r^{2}+\lambda^{2})}\right\} e^{3}_{i} \Bigg],  
%\\& (\theta = [ \tan^{-1}\frac{x_{0}}{\sqrt{r^{2}+\lambda^{2}}} + (n + \frac{1}{2})\pi ] \frac{r}{\sqrt{r^{2}+\lambda^{2}}}),   
 \end{split}
\end{equation}
where we have introduced $e^{1,2,3}_{i}$, which behave as three independent vectors under the combined rotation in the \lq\lq external\rq\rq 3D space $(x_{1}, x_{2}, x_{3})$ and in the \lq\lq internal\rq\rq gauge space of the adjoint representation $(\sigma_{1}, \sigma_{2}, \sigma_{3})$ \cite{Yaffe}:  
\be 
\label{1.9} 
e^{1}_{i} = \epsilon_{ijk}\sigma_{j}\hat{x}_{k}, \ \ e^{2}_{i} = (\delta_{ij} - \hat{x}_{i}\hat{x}_{j})\sigma_{j}, \ \ e^{3}_{i} = \hat{x}_{i} \hat{x}_{j}\sigma_{j}.  
\ee 

We are now ready to demonstrate that the time slice at $x_{0}= 0$ of this solution really carries $N_\CS  = n + \frac{1}{2}$ by an explicit calculation. 
Before that, we attempt to argue that this is also naturally expected from a general symmetry argument.
As stated above, the theory has a translational invariance concerning the change of $N_\CS $. Also, \EQREF{1.2} tells us $N_\CS $ is odd under the parity transformation, though the theory itself is parity invariant.
Thus the energy of the field configuration should be even function of $N_\CS $.
Suppose the extremum of the energy, namely the sphaleron-like configuration, is realized at $N_\CS  = n + a \ (0 < a <1)$, then the same energy is obtained for $N_\CS  = - n - a$ and $N_\CS  = n + 1 - a$ 
because of the parity and translational invariance.
This suggests that the sphaleron-like configuration is realized at $a = \frac{1}{2}$ as we expected, where $a = 1 - a$ holds.
On the other hand, the theory is also T-invariant.
So the energy should be even function of $x_{0}$ (as is also shown by an explicit calculation; see \EQREF{1.17}) and therefore the sphaleron-like configuration is expected to be realized at $x_{0} = 0$.
Thus our expectation is that $N_\CS  = n + \frac{1}{2}$ is achieved at the time slice $x_{0} = 0$. 

In order to demonstrate that this is really the case, what we need is 
$A'_{i}|_{x_{0} = 0} \equiv A'_{i}(0)$ and $F'_{0i}(0) \equiv \partial_{0} A'_{i}|_{x_{0} = 0}$ 
(equivalent to $\frac12 \epsilon_{ijk}F'_{jk}(0)$ under the self-duality).  
$A'_{i}(0)$ is given as 
\begin{equation}
\label{1.10}  
  A'_{i}(0)= i \left[ 
  \left\{\frac{1}{2r} + \frac{r^{2}-\lambda^{2}}{2r(r^{2}+\lambda^{2})} \cos 2\theta_{0}\right\} e^{1}_{i} 
- \frac{r^{2}-\lambda^{2}}{2r(r^{2}+\lambda^{2})} \sin 2\theta_{0} \ e^{2}_{i}  
+  \lambda^{2}\frac{\theta_{0}}{r(r^{2}+\lambda^{2})} \ e^{3}_{i} \right]
\end{equation}
with
\begin{equation}
  \theta_{0} =\left (n + \frac{1}{2}\right)\pi \frac{r}{\sqrt{r^{2}+\lambda^{2}}}.
\end{equation}
By use of $\partial_{0}\theta |_{x_{0}=0} = \frac{r}{r^{2}+\lambda^{2}}$, $F'_{0i}(0)$ is known to take the following simple form 
\be 
\label{1.11}
F'_{0i}(0) = 2i \frac{\lambda^{2}}{(r^{2}+\lambda^{2})^{2}} \left( \sin 2\theta_{0} \ e^{1}_{i} + \cos 2\theta_{0} \ e^{2}_{i} + e^{3}_{i} \right).   
\ee 

As a useful formula, when $A_{i}$ is written as $A_{i} = i (\alpha e^{1}_{i} + \beta e^{2}_{i} + \gamma e^{3}_{i})$ in general,  
\be 
\label{1.12}
\epsilon_{ijk} {\rm Tr} (A_{i}A_{j}A_{k}) = 12 (\alpha^{2}+\beta^{2})\gamma. 
\ee 
Though we skip the detail of the calculation, by use of this formula, one piece of $N_\CS $ is calculated to be  
\be 
\label{1.13} 
\frac{1}{24\pi^{2}} \int d^{3}x~ 
\epsilon_{ijk} {\rm Tr} \left\{A'_{i}(0)A'_{j}(0)A'_{k}(0)\right\} 
= \frac{11}{15}\left(n+\frac{1}{2}\right) - \frac{1}{\pi^{2}(n+\frac{1}{2})}.  
\ee 
We also get another piece (by use of ${\rm Tr}(e^{1}_{i}e^{1}_{i}) = {\rm Tr}(e^{2}_{i}e^{2}_{i}) = 4, \ {\rm Tr}(e^{3}_{i}e^{3}_{i}) = 2$, etc.): 
\begin{equation}
\label{1.14}
 \begin{split}
 - \frac{1}{16\pi^{2}} \int d^{3}x ~\epsilon_{ijk} {\rm Tr} \{ A'_{i}(0)F'_{jk}(0) \} 
  =& - \frac{1}{8\pi^{2}} \int d^{3}x~ {\rm Tr} \{ A'_{i}(0)F'_{0i}(0) \} 
 \\ 
  =& \frac{4}{15}\left(n+\frac{1}{2}\right) + \frac{1}{\pi^{2}(n+\frac{1}{2})}. 
 \end{split}
\end{equation}
Combining these two pieces, we obtain 
\be 
\label{1.15}
N_\CS  = n+\frac{1}{2}, 
\ee 
as we anticipated.  In particular, \EQREF{1.10} with $\theta_{0} = \frac{\pi}{2} \frac{r}{\sqrt{r^{2}+\lambda^{2}}}$, corresponding to  
$n = 0$, is a gauge field configuration with $N_\CS  = \frac{1}{2}$, just as the well-known sphaleron is. 

Unfortunately, however, this solution to the (Euclidean) equation of motion in ordinary 4D space-time faces an essential difficulty, when we regard it as the configuration to provide the potential barrier to path through under the circumstance of finite temperature $T$, where the transition probability between the vacua with $N_\CS  = n$ and $n+1$ is handled by a factor $e^{-\frac{M}{T}}$, with $M$ being the energy or \lq\lq mass\rq\rq possessed by the relevant field configuration. 

Namely, the energy possessed by the slice at $x_{0} = 0$ of the instanton solution, though it carries $N_\CS  = \frac{1}{2}$ (for $n = 0$), is not fixed but varies depending on the size of the instanton $\lambda$. In fact, $F_{0i}$ derived from \EQREF{1.1} is easily calculated to be 
\be 
\label{1.16}
F_{0i} = 2i \frac{\lambda^{2}}{(\rho^{2}+\lambda^{2})^{2}} (e^{2}_{i}+e^{3}_{i}),  
\ee 
which leads (with the help of self-duality) to the energy 
\be 
\label{1.17} 
E = 2\left(-\frac{1}{g^{2}}\right) \int d^{3}x \ {\rm Tr} (F_{0i})^{2} = \frac{6\pi^{2}}{g^{2}}\frac{\lambda^{4}}{(x_{0}^{2}+\lambda^{2})^{\frac{5}{2}}},  
\ee 
while the action of the instanton is easily seen, by the $x_{0}$-integral of this result, to be independent of $\lambda$: $\frac{8\pi^{2}}{g^{2}}$.
Thus for the slice $x_{0} = 0$, its energy is $\frac{6\pi^{2}}{g^{2}}\frac{1}{\lambda}$. Thus, the transition probability between distinct vacua is $\lambda$-dependent and ill-defined.
This problem may have its origin in the fact that for the slice at $x_{0} = 0$, the infinity of 3D space has a topology of S$^{2}$ and the relevant homotopy becomes trivial, i.e. 
$\pi_{2}({\rm SU(2)}) = 0$: the slice can be shrunk to a trivial vacuum with $A_{i} = 0$.

\section{The BPS monopole in the scenario of gauge-Higgs unification as sphaleron-like solution} 

The problem discussed in the previous section may be resolved invoking the 5D gauge theory, especially in the framework of gauge-Higgs unification (GHU), since now spatial infinity is topologically S$^{3}$ (not S$^{2}$) and therefore the mapping into the gauge manifold is non-trivial: $\pi_{3}({\rm SU}(2)) = {\bf Z}$ or $\pi_{3}({\rm S}^{2}) = {\bf Z}$ as in the case discussed in this section.
Also, in contrast to the case of 4D instanton solution, the extra space component of the gauge field $A_y$ may be regarded as an adjoint scalar field (Higgs fields) from 4D point of view, which makes difference in the analysis. 

In this section, by virtue of the characteristic feature of GHU, we can achieve analytic construction of a sphaleron-like configuration, i.e. a static finite energy solution to the equation of motion with $N_\CS  = \frac{1}{2}$.
The solution is actually a self-dual gauge field (in the sense of spatial 4D including the extra dimension with gauge fields $(\vec{A}, A_y)$, not ordinary 4D space-time).
It is nothing but the \lq\lq BPS saturated\rq\rq ~'t Hooft-Polyakov monopole (TPM) \cite{'t Hooft_monopole, Polyakov}, called BPS monopole, although the vacuum expectation value (VEV) of the adjoint scalar field and therefore the mass of the monopole, which is left as a free parameter and cannot be fixed in the original BPS monopole, is now found to be fixed and quantized in this higher dimensional framework as we will see below, just as the action is quantized in the case of ordinary 4D instanton.
Thus the factor to describe the transition probability at finite temperature between the distinct vacua is now well-defined.     

Our model is just 5D SU(2) GHU model.
For brevity, we ignore the potential term for $A_{y}$, induced at the quantum level.
This may be a reasonable assumption, as we are interested in the classical solution of the gauge field, and also because the BPS monopole is achieved in the limit of vanishing potential for the adjoint scalar.
The extra dimension is assumed to be a circle of radius $R$, whose coordinate is $y$ ($-\pi R \leq y \leq \pi R$).
The higher dimensional gauge field is $A_{M} = (A_{\mu}, A_{y}) \ (\mu = 0, 1, 2, 3)$, whose extra space component $A_y$ behaves as an adjoint scalar field from 4D point of view.
Namely, the adjoint scalar inevitably needed in the construction of the BPS monopole is already built-in.

The action of the model to start with is 
\be 
\label{2.1} 
S = \int d^{4}x dy ~\frac{1}{2g^{2}} {\rm Tr} \left(F_{MN}F^{MN}\right) \  \ \ \ (M,N = \mu, y \ (\mu = 0,1,2,3)),   
\ee 
where we again use the notation that the $A_{M}$ corresponds to $-ig A_{M}$ in the ordinary notation and accordingly $F_{MN} = \partial_{M}A_{N}-\partial_{N}A_{M}+[A_{M},A_{N}]$.
Let us note that the combination $g A_{M}$ does not depend on the space-time dimension: $gA_{M} = g_{4}A^{(4)}_{M}$, where $g_{4}, \ A^{(4)}_M$ are 4D gauge coupling constant and gauge fields, respectively.  

Since we are interested in a static sphaleron-like solution, we assume the gauge fields are all $x_{0}$-independent and $A_{0} = 0$ is understood.
Then $F_{0\mu}$ and $F_{0y}$ can be neglected and the issue is the minimization of the Hamiltonian of the system, which contains only the spatial components of the gauge field: 
\be 
\label{2.2} 
H = \int d^{3}x dy~{\cal H} \ ,
\ \ \ {\cal H} = - \frac{1}{2g^{2}}{\rm Tr}\left(F_{IJ}\right)^{2} 
= -\frac{1}{4g^{2}}{\rm Tr}
\left[ \left(F_{IJ} \pm \tilde{F}_{IJ}\right)^{2} \mp 2 F_{IJ}\tilde{F}_{IJ} \right], 
\ee 
where $\tilde{F}_{IJ} \equiv \frac{1}{2} \epsilon_{IJKL} F_{KL}$ ($I, J, K, L = 1, 2, 3, y$).
Just as in the case of the instanton in ordinary 4D space-time, for fixed winding number given by the integral of the term proportional to $F_{IJ}\tilde{F}_{IJ}$, the minimum of the energy is realized by the \lq\lq (anti-)self-dual\rq\rq solution in the (spatial) 4D space satisfying the self-duality condition: 
\be 
\label{2.3} 
F_{IJ} = \pm \tilde{F}_{IJ}. 
\ee 
Needless to say, such self-dual configuration of the gauge field is the solution to the equation of motion automatically.

By the way, \EQREF{2.3} suggests the presence of \lq\lq space-like instanton\rq\rq which is $y$-dependent and localizes in the 4D space (not 4D space-time), readily obtained by replacing $A_{0}$ and $x_{0}$ in \EQREF{1.1} by $A_{y}$ and $y$, respectively, as long as $\lambda$ is sufficiently small compared with $R$ \cite{HL}.
The relevant homotopy is $\pi_{3}({\rm SU}(2)) = {\bf Z}$ just as in the case of 4D instanton.    

Here, let us focus on the low-energy effective theory of this model by concentrating on the sector of $y$-independent Kaluza-Klein (KK) zero mode of the gauge field.
Then $F_{i y} = D_{i}A_{y} \ (D_{i}A_y = \partial_{i}A_{y} + [A_{i}, A_{y}])$, with $D_{i}A_{y}$ 
corresponding to the gauge covariant derivative of the adjoint \lq\lq scalar\rq\rq $A_{y}$ from the viewpoint of the 3D space.
Then the self-duality condition \EQREF{2.3} now reads as 
\be 
\label{2.4} 
F_{ij} = \pm \epsilon_{ijk} D_{k}A_{y}. 
\ee 
Interestingly, this turn out to be nothing but the BPS condition for the TPM, once $A_y$ is identified with an adjoint scalar \cite{HL}.
The solution of \EQREF{2.4} is readily obtained by utilizing the field configurations of the BPS monopole:  
\begin{equation}
\label{2.5}
 \begin{dcases}
 & A_{y} = -i F(r)\frac{gv}{2} \hat{x}_{i} \sigma_{i}, \\ 
& A_{i} = -i G(r) \frac{1}{r}\epsilon_{ijk}\hat{x}_{j}\frac{\sigma_{k}}{2}
  = i G(r)\frac{1}{2r} e^{1}_{i},  
 \end{dcases}
\end{equation}
with $F(\infty) = G(\infty) = 1$ and $F(0) = G(0) = 0$ as the boundary conditions.
The gauge coupling constant $g$ and the VEV $v$ may be understood to be those for 4D space-time, if we wish. 

Since the vacuum state of this system, realized at the spatial infinity $r \to \infty$, shows the \lq\lq hedgehog-type\rq\rq field configuration proportional to the factor  $\hat{x}_{i} \sigma_{i}$, with $\hat{x}_{i}$ being the coordinates to describe S$^{2}$, here 
the relevant homotopy is $\pi_{3}({\rm S}^{2}) = {\bf Z}$.    
The condition \EQREF{2.4} is equivalent to the following coupled differential equation 
\begin{equation}
\label{2.6} 
 \begin{dcases}
& gv F(1 - G) =  G' \\  
& gvr^{2} F' = G(2 - G) 
 \end{dcases}
,~~~~~~~~\left(F' = \frac{dF}{dr}, \ G' = \frac{dG}{dr}\right)
\end{equation}
whose solution is well-known:     
\begin{equation}
\label{2.7} 
 \begin{dcases}
& F(r) =  \coth (gvr) - \frac{1}{gvr},  \\  
& G(r) =  1 - \frac{gvr}{\sinh (gvr)}. 
 \end{dcases}
\end{equation}

Although these field configurations themselves belong to the $y$-independent KK zero modes of the fields, $y$-dependence is expected to arise when we perform a $y$-dependent local gauge transformation in order to eliminate $A_y$: $A_{y} \to A'_{y} = 0$.
This is because the parameter for the gauge transformation should be linear function of $y$, not periodic in $y$, since the KK zero mode of $A_y$ itself is $y$-independent.
From the lesson we learnt in the previous section for the example of 4D instanton, it would be reasonable to expect that, after the gauge transformation, the slice at $y = 0$ of $A_{i}$ has $N_\CS  = n + \frac{1}{2}$, with $n$ being an integer.

We now show that this really is the case by use of concrete calculations.
We first perform a gauge transformation $A_{y} \to A'_{y} = V^{-1}A_{y}V + V^{-1}\partial_{y} V$ 
due to an unitary matrix $V = e^{i \theta \hat{x}_{i}\cdot \sigma_{i}}$.
It is easy to see that $A'_{y} = 0$ is realized by choosing 
\be 
\label{3.1} 
\theta = F(r) \frac{gv}{2}\left\{ y + \left(n + \frac{1}{2}\right)2\pi R\right\}, 
\ee 
where the constant of integration $(n + \frac{1}{2})2\pi R$ has been chosen so that the CS number changes as $N_\CS  = n \to n+1$ for $y = -\pi R \to \pi R$, as is shown below. 

Though we skip the details of the calculation, the transformed gauge field $A'_{i} = V^{-1}A_{i}V + V^{-1}\partial_{i} V$ is obtained by a similar calculation to the one performed in the previous  section for the case of the instanton: 
\begin{equation}
\label{3.2} 
 \begin{split}
  A'_{i} =& i \left[ \frac{(G-1)\cos 2\theta + 1}{2r} e^{1}_{i} 
  - \frac{(G-1)\sin 2\theta }{2r}e^{2}_{i} 
  + F' \frac{gv}{2}\left\{y + \left(n + \frac{1}{2}\right)2\pi R \right\} e^{3}_{i} \right]  \\ 
  =& i \Bigg[ 
  \frac{1}{2}\left\{- \frac{gv}{\sinh (gvr)} \cos 2\theta + \frac{1}{r}\right\} e^{1}_{i} 
  + \frac{1}{2}\frac{gv}{\sinh (gvr)}\sin 2\theta ~e^{2}_{i}
  \\ 
  &~~~~~~~+ \left\{-(gv)^{2} \frac{1}{\sinh^{2}(gvr)} + \frac{1}{r^2}\right\} 
  \frac{1}{2}\left\{y + \left(n + \frac{1}{2}\right)2\pi R \right\} e^{3}_{i} \Bigg],
 \end{split}
\end{equation}
where 
\be 
\label{3.3} 
\frac{d\theta}{dr} = F' \frac{gv}{2}\left\{y + \left(n + \frac{1}{2}\right)2\pi R \right\}. 
\ee
We will also need 
\be 
\label{3.4} 
F'_{yi} = \frac{\partial A'_{i}}{\partial y} 
= i \left[ -\frac{gv}{2} \frac{F(G-1)}{r}(\sin 2\theta \ e^{1}_{i} + \cos 2\theta \ e^{2}_{i}) + \frac{gv}{2}F' \ e^{3}_{i} \right].  
\ee

\subsection{The Chern-Simons number}  

We are now ready to calculate the CS number $N_\CS $: 
\be 
\label{3.5}  
N_\CS  = -\frac{1}{16\pi^{2}} \int d^{3}x \ \epsilon _{ijk} \ {\rm Tr}
\left[A'_{i}F'_{jk} - \frac{2}{3}A'_{i}A'_{j}A'_{k} \right]. 
\ee
One piece of $N_\CS $ is (referring to \EQREF{1.12})
\begin{equation}
\label{3.9} 
 \begin{split}
 &\frac{1}{24\pi^{2}}\int d^{3}x \ \epsilon _{ijk} \ {\rm Tr} 
\left(A'_{i}A'_{j}A'_{k}\right) \\ 
  =& \frac{1}{2\pi} \int_{0}^{\infty} dr \ 
  \left[ (G-1)^{2} + 1 + 2(G-1)\cos 2\theta \right]\cdot \frac{d\theta}{dr},
 \end{split}
\end{equation}
and another one is  
\begin{equation}
\label{3.10}
 \begin{split}
& - \frac{1}{16\pi^{2}}\int d^{3}x \ \epsilon _{ijk} \ 
  {\rm Tr}\left(A'_{i}F'_{jk}\right) 
  = - \frac{1}{8\pi^{2}} \int d^{3}x \ {\rm Tr}\left(A'_{i}F'_{yi}\right)\\ 
  =& - \frac{1}{2\pi} \int_{0}^{\infty} dr \  
   \left[ gv F(G-1) \sin 2\theta - gvr^{2} F' \frac{d\theta}{dr} \right]\\ 
  =& \frac{1}{2\pi} \int_{0}^{\infty} dr \  
  \left[ G'\sin 2\theta +G(2-G) \frac{d\theta}{dr} \right],  
 \end{split}
\end{equation}
where in the first line we relied on the self-duality of the gauge field and also the relations \EQREF{2.6} have been used in the third line. 
Summing up these two pieces the integrand turns out to be a total derivative, and by use of $F(0) = 0, \ F(\infty) = G(\infty) = 1$ and \EQREF{3.1}, 
\begin{equation}
\label{3.12}
 \begin{split}
 N_\CS  =& \frac{1}{2\pi} \int_{0}^{\infty} dr 
  \frac{d}{dr}\left[(G-1) \sin 2\theta + 2\theta\right] \\ 
  =& \frac{1}{2\pi} gv \ \left[y + \left(n + \frac{1}{2}\right)2\pi R\right]. 
 \end{split}
\end{equation}

For $y = - \pi R, \ 0, \ \pi R$, the CS number varies as $N_\CS  = (gvR)n, \ (gvR)(n + \frac{1}{2}), \ (gvR)(n+1)$, similarly to the change of the $N_\CS $ for $t = - \infty, \ 0, \ 
\infty$ in the case of ordinary 4D instanton.
Hence, for $n = 0$, the slice of this solution at $y = 0$ may be regarded as the sphaleron-like solution, which carries $N_\CS  = \frac{1}{2}$, once the VEV of $A_{y}$ satisfies 
\be 
\label{3.14} 
v = \frac{1}{gR}. 
\ee 
In the next subsection, we argue that in fact this condition is realized by the topological nature of this self-dual gauge field. 

We thus conclude that we have achieved the construction of the following analytic sphaleron-like solution $A^\SP _{i}(x)$, by setting $y = 0$ (also $n = 0$) and $v = 1/(gR)$ in \EQREF{3.2}          
\begin{equation}
\label{3.15} 
 \begin{split}
A^\SP _{i} =& i \frac{1}{2R} 
  \Bigg[ \left\{- \frac{1}{\sinh \left(\frac{r}{R}\right)}\cos 2\theta_{0} + \frac{R}{r}\right\} \ e^{1}_{i} 
  + \frac{1}{\sinh \left(\frac{r}{R}\right)}\sin 2\theta_{0} \ e^{2}_{i} \\ 
 &\ \ \ \ \ \ \ \  
  +\pi \left\{- \frac{1}{\sinh ^{2}\left(\frac{r}{R}\right)} + \left(\frac{R}{r}\right)^{2}\right\} \ e^{3}_{i} \Bigg], 
 \end{split}
\end{equation}
with 
\be 
\label{3.16} 
\theta_{0} = \frac{\pi}{2}F 
= \frac{\pi}{2}\left[\coth\left(\frac{r}{R}\right) - \frac{R}{r}\right]. 
\ee 
This solution has $N_\CS  = \frac{1}{2}$ independently of $R$. 

So far we have been working in the gauge, in which $A'_{y}$ vanishes.
Let us note that in this gauge, though $A'_{y}$ vanishes, the $y$-dependence of $A'_{i}$ should be taken into account properly in order 
to evaluate the contribution of $F'_{yi} = \partial_{y}A'_{i}$ (see \EQREF{3.4}) to the energy (the mass) of this sphaleron-like configuration properly.
Thus $A'_{i}$ no longer belongs to the KK zero mode.

However, once we have learnt how to construct the 4D gauge field with 
$N_\CS  = \frac{1}{2}$, we do not have to insist on this gauge 
and we can take another one, in which 4D gauge field and 4D adjoint scalar coexist 
and both are $y$-independent, behaving as KK zero modes.
We first realize that
\EQREF{3.15} is achieved by a gauge transformation from the original field $A_{i}$ in \EQREF{2.5} 
by an unitary matrix $V_{0} = e^{i \theta_{0} \hat{x}_{i}\cdot \sigma_{i}}$, 
where $\theta$ in \EQREF{3.1} is replaced from the beginning by $\theta_{0}$ given in \EQREF{3.16}. 
So, we now consider the gauge transformation from the original fields $A_{i}$ and $A_{y}$ in \EQREF{2.5} 
by the unitary matrix $V_{0}$. 
As is stated above, $A_{i}$ is transformed into $A^\SP _{i}$ given in \EQREF{3.15}. 
On the other hand, concerning the 4D scalar field, now the transformed field 
$A'_{y}$ does not vanish, since $\theta_{0}$ is $y$-independent, 
and actually is the same as the original field, as is readily known from 
$[V_{0}, A_{y}] = 0$. 
In this way, we are able to construct $y$-independent sphaleron-like field configurations:  
\begin{equation}
\label{3.17}
 \begin{split}
 A^\SP _{i} =& i \frac{1}{2R} 
  \Bigg[ \left\{- \frac{1}{\sinh (\frac{r}{R})}\cos 2\theta_{0} + \frac{R}{r}\right\} \ e^{1}_{i} 
  + \frac{1}{\sinh (\frac{r}{R})}\sin 2\theta_{0} \ e^{2}_{i} 
  \\
  &\ \ \ \ \ \ \ \ 
  + \pi \left\{- \frac{1}{\sinh ^{2}(\frac{r}{R})} + \left(\frac{R}{r}\right)^{2}\right\} \ e^{3}_{i}
  \Bigg], \ \ \ 
  %(\theta_{0} = \frac{\pi}{2}\left[\coth(\frac{r}{R}) - \frac{R}{r}\right]),  
  \\ 
A^\SP _{y} =& A_{y} = -i \frac{1}{2R} 
  \left[ \coth \left(\frac{r}{R}\right) - \frac{R}{r}\right] \hat{x}_{i} \sigma_{i}.
 \end{split}
\end{equation}
Since $A^\SP _{i}$ is exactly the same as the one in \EQREF{3.15}, it of course carries $N_\CS  = \frac{1}{2}$.
Also the mass of this configuration, i.e. $M_\SP $, does not depend on the gauge choice (see \EQREF{3.18b}).

\subsection{Topologically fixed vacuum expectation value and the mass of the sphaleron-like solution}    

In the framework of the ordinary TPM, the strength of the VEV of the adjoint scalar, $v$, and therefore the mass of the BPS monopole is left as an arbitrary parameter in the BPS limit.
In our discussion based on GHU, however, the sphaleron-like solution may be interpreted as a self-dual gauge field in the 4D space, just as the instanton solution is in the 4D space-time.
Thus the sphaleron-like solution should carry the winding number 1, and therefore we naturally expect that $v$ and the mass possessed by this solution is topologically quantized.

Let us recall that the Hamiltonian of our system is given as (see \EQREF{2.2}) 
\be 
\label{3.17a} 
H = \int d^{3}x dy~{\cal H} \ , 
\ \ \ {\cal H} = - \frac{1}{2g^{2}}{\rm Tr}(F_{IJ})^{2},      
\ee 
where $- \int d^{3}x dy \ \frac{1}{2}{\rm Tr}(F_{IJ})^{2}$ should be quantized (with a winding number $\nu$), just as in the case of instanton:   
\be  
\label{3.18} 
- \int d^{3}x dy \ \frac{1}{2}{\rm Tr}\left(F_{IJ}\right)^{2} = 8\pi^{2} \nu. 
\ee 
Thus the energy of the topological configuration is now quantized as 
\be 
\label{3.18a} 
E = \frac{8\pi^{2}}{g^{2}}\nu = \frac{4\pi}{g_{4}^{2}}\frac{1}{R}\nu,  
\ee 
where $g_{4}$ is the 4D gauge coupling constant. For $\nu = 1$, this gives the mass of our sphaleron-like solution: 
\be 
\label{3.18b} 
M_\SP  = \frac{8\pi^{2}}{g^{2}} = \frac{4\pi}{g_{4}^{2}}\frac{1}{R}. 
\ee
On the other hand, it is well-known that the mass of the BPS monopole is given by $\frac{4\pi v_{4}}{g_{4}}$ ($v_{4}$ denotes the VEV of the 4D field).
Thus, comparing with \EQREF{3.18b}, we conclude 
\be 
\label{3.18c} 
v_{4} = \frac{1}{g_{4}R}, \ \ \ {\rm or} \ \ v =\frac{1}{gR},   
\ee 
as we expected.
It is interesting to note that the $v$ determined by the radiatively induced potential for $A_y$ is just $\frac{1}{gR}$ for SU(2) GHU model, a toy model for electro-weak unified GHU \cite{Kubo}.

The mass $M_\SP $ given in \EQREF{3.18b} is responsible for the Boltzmann factor to describe the transition probability between distinct vacua under the circumstance of finite temperature.
If we naively assume that $g_{4}v_{4} \sim  \frac{1}{R} \sim M_{W}$ as in the electro-weak unified GHU, we can make an estimation 
\be 
\label{3.22} 
M_\SP  \sim \frac{M_{W}}{\alpha}  ~~~~~~~~ \left(\alpha = \frac{e^{2}}{4\pi}\right),    
\ee 
which is of $\cal{O}$(10) TeV.

\section{On the instability of the sphaleron-like solution}

The sphaleron is a saddle point of the energy in the space of gauge field configurations.
Namely, the configuration is unstable and the energy decreases when the field configuration gradually deforms along the direction, in which the $N_\CS $ varies.
In this section, we address the question whether the sphaleron-like solution we constructed in the previous section also may be regarded as a saddle point. 

First let us note the importance of the boundary condition of the fields $A_{i}$ and $A_{y}$ in this context of the argument.
In the manipulation of \EQREF{2.2}, 
the term $F_{IJ}\tilde{F}_{IJ}$ in the Hamiltonian density is well-known to be written as a total derivative.
Thus its contribution to the energy $E$ of the system is given just by the boundary conditions of the fields at spatial infinity: $A_{i}(r = \infty), \ A_{y}(r = \infty)$.
Since the remaining term in the Hamiltonian density is semi-positive-definite, we easily find that any deformation of the fields from the sphaleron-like solution \EQREF{3.17}, for which the remaining term vanishes because of the self-duality, will only increase the energy $E$.
Namely, Hessian at the point of the sphaleron-like solution should not have negative eigenvalue, as long as the boundary condition is fixed. 

This situation may be understood concretely, by plugging \EQREF{2.5} into \EQREF{2.2}, to get 
\begin{equation}
\label{4.1} 
 \begin{split}
 E =& \frac{4\pi^{2}R}{g^{2}} \int_{0}^{\infty} dr \ 
  \left[ 2G'^{2}+\frac{G^{2}(2-G)^{2}}{r^{2}} 
  + g^{2}v^{2}\left(r^{2}F'^{2}+2F^{2}(1-G)^{2}\right) \right]
  \\ 
=& \frac{4\pi^{2}R}{g^{2}} \int_{0}^{\infty} dr \ 
  \bigg[ 2\left(G' - gv(1-G)F\right)^{2}  \\ 
   &~~~~~~~~~~~~~~~~~~~~~~
   + \left(\frac{G(2-G)}{r} - gvrF'\right)^{2}
   + 2gv \frac{d}{dr}\left(G(2-G)F\right) \bigg]. 
 \end{split}
\end{equation}
In this calculation, we have adopted the original fields given in \EQREF{2.5} instead of the gauge transformed fields $A'_{i, y}$, since the energy is gauge invariant.
Under the BPS (or self-duality) condition, \EQREF{2.6}, only the last term of \EQREF{4.1} contributes to the energy, i.e. the mass of the sphaleron-like solution  
$M_\SP  = \frac{8\pi^{2}Rv}{g} = \frac{4\pi v_{4}}{g_{4}}$, with the boundary conditions $F(\infty) = G(\infty) = 1, \ F(0)= G(0)= 
0$ being taken into account.
Thus, as long as the boundary conditions $A_{i}(r = \infty), \ A_{y}(r = \infty)$ are fixed, any deformation of $F, \ G$ from the sphaleron-like solution, \EQREF{3.17}, does not reduce the energy. 

Hence, assuming we keep the hedgehog-type field configuration as is shown in \EQREF{2.5}, the only possibility to reduce the energy will be to modify the boundary conditions of $A_{i}, \ A_{y}$ at spatial infinity $r \to \infty$, especially that of $A_{y}$.
Probably, the most simple example of the boundary condition's deformation is to replace the VEV $v$ by a parameter-dependent quantity $v(\mu)$ with  $\mu$ being the parameter.
One remark here is that in the scenario of GHU, as was mentioned in the introduction, the potential of $A_{y}$ (to be understood as a Higgs field), $V(A_y)$, is a periodic function with a period $\frac{2}{gR}$, since $A_y$ has a physical meaning as the phase of Wilson-line.
We also note that $V(A_y)$ 
is an even function of $A_y$, since our model is parity symmetric and $A_y$ behaves as a pseudo-scalar field.
Hence, for a constant $a \ (0 \leq a \leq \frac{1}{gR})$, 
\be 
\label{4.1'}
V\left(\frac{1}{gR}+a\right) = V\left(-\frac{1}{gR}-a\right) 
= V\left(\frac{1}{gR}-a\right).
\ee 
This means, together with the periodicity, $V\left(A_{y} + \frac{2}{gR}\right) = V(A_{y})$, that the VEV $v$ may be restricted to the basic region, $0 \leq v \leq \frac{1}{gR}$.

Taking into account the presence of such basic region, we choose the following parameterization    
\be 
\label{4.2} 
v(\mu) = v\cos \mu \  \ \ \ \left(-\frac{\pi}{2} \leq \mu \leq \frac{\pi}{2}\right),  
\ee 
so that $v(0) = v = \frac{1}{gR}, \ v\left(\pm \frac{\pi}{2}\right) = 0$: the parameter $\mu$ connects the sphaleron-like solution with a trivial vacuum, $A_{i} = A_{y} = 0$ realized at $\mu = \pm \frac{\pi}{2}$.
Then what we should do is just to replace $v$ by $v(\mu)$ in \EQREF{2.5} and \EQREF{2.7}.
Since in the ordinary 4D space-time the BPS condition, \EQREF{2.6}, holds irrespectively of the specific value of $v$, such obtained field configurations still satisfy the condition and therefore equations of motion as well.
Now the energy of the system is modified into 
\be 
\label{4.3}
E = \frac{8\pi^{2}Rv(\mu)}{g} =  \ M_\SP\cos \mu  = \cos \mu \ \frac{4\pi}{g_{4}^{2}}\frac{1}{R}, 
\ee 
which reaches its maximal value $M_\SP $ at $\mu = 0$ and decreases for $\mu \neq 0$: such $\mu$-dependent deformation of $A_{i}, \ A_{y}$ from the sphaleron-like solution, corresponding to $\mu = 0$, is confirmed to reduce the energy.
It is interesting to note that the $N_\CS $ also changes depending on the parameter $\mu$, as is seen from \EQREF{3.12} (for $n = y = 0$):  
\be 
\label{4.4}
N_\CS  = \frac{1}{2}gv(\mu) R = \frac{1}{2} \cos \mu.  
\ee 
Thus, the sphaleron-like solution may be regarded as a saddle point, where the energy decreases according to the change of $N_\CS $, as we anticipated, while it gives the minimum energy for a fixed winding number.

The situation may have some similarity to the case of the instanton in ordinary 4D space-time, where the energy possessed by the instanton solution takes its maximal at $t = 0$ and also $N_\CS $ deviates from $\frac{1}{2}$ for $t \neq 0$, as we have already seen, though we do not intend to regard the parameter $\mu$ as related to $t$, since we are interested in static solutions and also because the equations of motion for gauge fields, now including $A_{0}$ as well, should be modified if time dependence is introduced. 

In the example of field deformation discussed above, at $\mu = \pm \frac{\pi}{2}$ all fields just vanish: $A_{i} = A_{y} = 0$, while what we expect as a trivial vacuum state is vanishing $A_{i}$ and non-vanishing constant $A_{y}$ satisfying 
$\frac{1}{2}{\rm Tr}A_{y}^{2} = -\left(\frac{gv}{2}\right)^{2}$, say $A_{y} = -i \frac{gv}{2}\sigma_{3}$ (which may be assumed without loss of generality thanks to the global SU(2) gauge symmetry), rather than $A_{y} = 0$.

Therefore, we next attempt to deform the fields, not imposing the hedgehog type field configuration, so that the deformation modifies the sphaleron-like solution with $N_\CS  = \frac{1}{2}$ into the trivial vacuum $A_{i} = 0, \ A_{y} = -i \frac{gv}{2}\sigma_{3}$ with $N_\CS  = 0$.
To be specific, we consider a parameter-dependent deformation of the original field (before the gauge transformation) of $A_{y}$ given in \EQREF{2.5}, by replacing the zenith angle $\theta$ in the spherical coordinate system 
$(r, \theta, \varphi)$ by $\tilde{\theta} =  \theta\cos\mu $ and
$r$ by $\tilde{r} = \frac{r}{\cos \mu}$ ($-\frac{\pi}{2} \leq \mu \leq \frac{\pi}{2}$):  
\be 
\label{4.5}  
\tilde{A}_{y} = -i\frac{gv}{2}F(\tilde{r}) (\sin \tilde{\theta}\cos \varphi \ \sigma_{1} + \sin \tilde{\theta}\sin \varphi \ \sigma_{2} + \cos \tilde{\theta} \ \sigma_{3}),   
\ee 
which recovers the BPS monopole solution for $\mu = 0$, while the case $\mu = \pm \frac{\pi}{2}$ corresponds to the trivial vacuum $\tilde{A}_{y} = -i \frac{gv}{2}\sigma_{3}$, since $\tilde{\theta} \to 0$ and $\tilde{r} \to \infty$
 ($F(\infty) = 1$) for $\mu \to \pm \frac{\pi}{2}$. 

In order to determine the deformed vector potential $\tilde{A}_{i}$, we first discuss the behavior of the fields at spatial infinity $r = \infty$, $\tilde{A}_{y}(\infty)$ and $\tilde{A}_{i}(\infty)$, and impose the condition that they satisfy $\tilde{F}_{iy} = \tilde{D}_{i} \tilde{A}_{y} = \partial_{i} \tilde{A}_{y} + [\tilde{A}_{i}, \tilde{A}_{y}] = 0$, with $\tilde{D}_{i}$ being understood as gauge covariant derivative for the adjoint \lq\lq scalar\rq\rq $\tilde{A}_{y}$.      
In the spherical coordinate system, the nabla vector is well-known to be written as  
\be 
\label{4.6} 
\vec{\nabla} = \vec{e}_{r} \frac{\partial}{\partial r}+ \vec{e}_{\theta} \frac{1}{r}\frac{\partial}{\partial \theta} 
+ \vec{e}_{\varphi} \frac{1}{r\sin \theta}\frac{\partial}{\partial \varphi},   
\ee 
where $\vec{e}_{r} = \vec{\hat{x}}, \ \vec{e}_{\theta}, \ \vec{e}_{\varphi}$ are unit vectors in the directions of $r, \ \theta, \ \varphi$, respectively.
Accordingly, the vector potential of the BPS monopole solution $\vec{A}$ at spatial infinity, for instance, can be written as 
\begin{subequations}
\label{4.7}  
\begin{equation}
 \vec{A}(\infty) = \vec{e}_{r} A_{r}(\infty)+ \vec{e}_{\theta} \frac{1}{r}A_{\theta}(\infty)+ \vec{e}_{\varphi} \frac{1}{r\sin \theta}A_{\varphi}(\infty),  
\end{equation}
where
\begin{equation}
  \begin{dcases}
   & A_{r}(\infty) = 0, \\
   & A_{\theta}(\infty) = \frac{i}{2}(-\sin \varphi \ \sigma_{1} + \cos \varphi \ \sigma_{2}), 
  \\ 
   & A_{\varphi}(\infty) = \frac{i}{2}\sin \theta (-\cos \theta \cos \varphi \ \sigma_{1}-\cos \theta \sin \varphi \ \sigma_{2} 
+ \sin \theta \ \sigma_{3}). 
  \end{dcases}
\end{equation}
\end{subequations}
$\vec{\tilde{A}}$ is basically obtained by replacing $\theta$ in $\vec{A}$ by $\tilde{\theta}$.
In addition, by noting 
$\frac{\partial}{\partial \theta} = \cos \mu \frac{\partial}{\partial \tilde{\theta}}$, we put additional factor $\cos \mu$ for $\tilde{A}_{\theta}$: 
\begin{subequations}
\label{4.8}  
\begin{equation}
 \vec{\tilde{A}} = \vec{e}_{\theta} \frac{1}{r}\tilde{A}_{\theta} + \vec{e}_{\varphi} \frac{1}{r\sin \theta}\tilde{A}_{\varphi},   
\end{equation}
where
\begin{equation}
 \begin{dcases}
  &\tilde{A}_{\theta} = \frac{i}{2}G(\tilde{r}) \cos \mu \ (-\sin \varphi \ \sigma_{1} + \cos \varphi \ \sigma_{2}), 
  \\
  &\tilde{A}_{\varphi} = \frac{i}{2} G(\tilde{r}) \sin \tilde{\theta} (-\cos \tilde{\theta} \cos \varphi \ \sigma_{1}-\cos \tilde{\theta} \sin \varphi \ \sigma_{2} 
+ \sin \tilde{\theta} \ \sigma_{3}) .  
 \end{dcases}
\end{equation}
\end{subequations}
It is easy to see that $\vec{\tilde{A}}$ vanishes for $\mu = \pm \frac{\pi}{2}$.
It is also confirmed that 
$\partial_{\theta}\tilde{A}_{y}(\infty) + \left[\tilde{A}_{\theta}(\infty), \tilde{A}_{y}(\infty)\right] = 0, \ 
\partial_{\varphi}\tilde{A}_{y}(\infty) + \left[\tilde{A}_{\varphi}(\infty), \tilde{A}_{y}(\infty)\right] = 0$.
Thus $\tilde{F}_{iy} = \partial_{i}\tilde{A}_{y} + \left[\tilde{A}_{i}, \tilde{A}_{y}\right]$ 
vanishes as $r \to \infty$, and therefore the calculated energy possessed by these modified fields should be finite. 

In order to obtain the energy, we first calculate, say the \lq\lq electric field\rq\rq, $\tilde{E}_{i} 
\equiv - \tilde{F}_{iy}$: 
\begin{equation}
\label{4.9}
 \begin{split}
 \vec{\tilde{E}} =&  - \vec{\nabla}\tilde{A}_{y} - \left[\vec{\tilde{A}},\tilde{A}_{y}\right]  \\ 
  =& i\frac{gv}{2} \bigg[ \frac{1}{\cos \mu}F' (\sin \tilde{\theta}\cos \varphi \ \sigma_{1} + \sin \tilde{\theta}\sin \varphi \ \sigma_{2} + \cos \tilde{\theta} \ \sigma_{3}) \ \vec{e}_{r} \\ 
 &\ \ \ \ \ \ 
  + \cos \mu \ \frac{F(1-G)}{r} (\cos \tilde{\theta}\cos \varphi \ \sigma_{1}+\cos \tilde{\theta}\sin \varphi \ \sigma_{2} - \sin \tilde{\theta} \ \sigma_{3}) \ \vec{e}_{\theta} \\ 
 &\ \ \ \ \ \ 
  + \frac{F(1-G)}{r} \frac{\sin \tilde{\theta}}{\sin \theta}(-\sin \varphi \ \sigma_{1}+\cos \varphi \ \sigma_{2}) \ \vec{e}_{\varphi} \ 
\bigg], 
 \end{split}
\end{equation}
where $F'$ stands for $F'(\tilde{r}) \equiv \frac{dF(\tilde{r})}{d\tilde{r}}$, not $\frac{dF(\tilde{r})}{dr}$. 
As expected, for $r \to \infty$, $\vec{\tilde{E}}$ vanishes, independently of the value of $\cos \mu$, since $G(\infty) = 1, \ F'(\infty) = 0$.
Next the \lq\lq magnetic field\rq\rq $\tilde{B}_{i} \equiv \frac{1}{2}\epsilon_{ijk}\tilde{F}_{jk}$ is calculated (by use of the vector analysis in the spherical coordinate system) to be 
\begin{equation}
\label{4.10}
 \begin{split}
 \vec{\tilde{B}} =&  \vec{\nabla} \times \vec{\tilde{A}} + \left[~\vec{\tilde{A}},\vec{\tilde{A}}~\right] \\ 
  =& \frac{i}{2}\bigg[ 
  \cos \mu \frac{G(2-G)}{r^{2}}\frac{\sin \tilde{\theta}}{\sin \theta} 
  (\sin \tilde{\theta}\cos \varphi \ \sigma_{1}
  +\sin \tilde{\theta}\sin \varphi \ \sigma_{2}+ \cos \tilde{\theta} \ 
  \sigma_{3}) \ \vec{e}_{r}\\ 
  &\ \ \ \ \ \ \ 
  + \frac{1}{\cos \mu} \frac{G'}{r} \frac{\sin \tilde{\theta}}{\sin \theta} (\cos \tilde{\theta}\cos \varphi \ \sigma_{1}
+\cos \tilde{\theta}\sin \varphi \ \sigma_{2} - \sin \tilde{\theta} \ \sigma_{3}) \  \vec{e}_{\theta}  \nonumber \\ 
 &\ \ \ \ \ \ \ 
  + \frac{G'}{r}(-\sin \varphi \ \sigma_{1}+\cos \varphi \ \sigma_{2}) \ \vec{e}_{\varphi} 
\bigg]. 
 \end{split}
\end{equation}

Now the energy possessed by these field configurations is given as 
\be 
\label{4.11}
E = -\frac{2\pi R}{g^{2}} {\rm Tr} \left[ \int d^{3}x (\vec{\tilde{E}} - \vec{\tilde{B}})^{2} -2 \lim_{r \to \infty} 
r^{2}\int \sin \theta \ d\theta d\varphi \ \tilde{A}_{y}\vec{\tilde{B}}\cdot \vec{e}_{r} \right], 
\ee 
where we have used a relation 
${\rm Tr} \left(\vec{\tilde{E}}\cdot \vec{\tilde{B}}\right) 
= -{\rm Tr}\left[(\vec{\tilde{D}}\tilde{A}_{y}) \cdot \vec{\tilde{B}}\right] 
= -{\rm Tr} \left[\vec{\tilde{D}}\cdot \left(\tilde{A}_{y}\vec{\tilde{B}}\right)\right] 
= -\vec{\nabla}\cdot {\rm Tr} \left(\tilde{A}_{y}\vec{\tilde{B}}\right)$
due to the Bianchi identity and the Gauss's theorem, with the normal direction being extracted as $\tilde{A}_{y}\tilde{B}_{n} = \tilde{A}_{y}\vec{\tilde{B}}\cdot \vec{e}_{r}$.
Plugging \EQREF{4.5}, \EQREF{4.9} and \EQREF{4.10} into \EQREF{4.11} and by use of the relation \EQREF{2.6} (with $r$ being replaced by $\tilde{r}$) and $F(\infty) = G(\infty) = 1$, 
\begin{equation}
\label{4.12}
 \begin{split}
 E =& \frac{4\pi^{2}R}{g^{2}} \ \int_{0}^{\infty} dr \int_{0}^{\frac{\pi}{2}} \sin \theta \ d\theta \ 
  \Bigg[ \cos^{2}\mu \left(1-\frac{\sin \tilde{\theta}}{\sin \theta}\right)^{2} \frac{G^{2}(2-G)^{2}}{r^{2}} 
\\
  &\ \ \ \ \ \ \ \ 
  +  g^{2}v^{2}\left\{\left(\frac{1}{\cos \mu}\frac{\sin \tilde{\theta}}{\sin \theta} - \cos \mu\right)^{2} 
+\left(1-\frac{\sin \tilde{\theta}}{\sin \theta}\right)^{2} \right\} F^{2}(1-G)^{2}\Bigg] \\ 
&+ \frac{8\pi^{2}R}{g}v\left[1-\cos\left(\frac{\pi}{2} \cos \mu\right)\right], 
 \end{split}
\end{equation}
where the integral is over the \lq\lq northern hemisphere\rq\rq ($0 \leq \theta < \frac{\pi}{2}$) and then doubled, invoking the parity symmetry of the theory.
As we expected, the integral is finite, since $F(\infty) = G(\infty) = 1$, and we have succeeded in obtaining the finite energy deformation.
Also, for $\mu = 0$, $E$ just coincides with $M_\SP  = \frac{8\pi^{2}Rv}{g}$, while for $\mu \to \pm \frac{\pi}{2}$, $E$ vanishes, 
since $\sin \tilde{\theta} \simeq \tilde \theta =\theta \cos \mu $ and $G(\tilde{r}) = G(\infty) = 1$ in this limit.

Interestingly, without performing the integral explicitly in \EQREF{4.12}, we can easily confirm 
\be 
\label{4.13} 
\left. \frac{\partial E}{\partial \mu}\right|_{\mu = 0} = 0, \ \ 
\left. \frac{\partial^{2}E}{\partial \mu^{2}}\right|_{\mu = 0} 
= - \frac{4\pi^{3}Rv}{g} < 0,      
\ee 
which implies that the sphaleron-like solution at least gives local maximum of the energy.  
This argument can be further extended by rewriting \EQREF{4.12} as the sum of $M_\SP$ and the remaining terms. Namely, rewriting a piece of the surface term, $- \frac{8\pi^{2}R}{g}v \cos(\frac{\pi}{2} \cos \mu)$, in a form of the integral in the 4D bulk space, we obtain 
\begin{equation}
\label{4.13'}
 \begin{split}
E =& M_\SP  
\\ 
& 
+ \frac{4\pi^{2}R}{g^{2}} \ \int_{0}^{\infty} dr \int_{0}^{\frac{\pi}{2}} \sin \theta \ d\theta \ 
\Bigg[ \left\{ \cos^{2}\mu \left(1+\frac{\sin^{2} \tilde{\theta}}{\sin^{2} \theta} \right)
-2\cos \mu \right\} \frac{G^{2}(2-G)^{2}}{r^{2}} 
\\ 
&
+ g^{2}v^{2}\left\{ 1 + \cos^{2}\mu + \frac{\sin^{2} \tilde{\theta}}{\sin^{2} \theta} 
\left(1+ \frac{1}{\cos^{2}\mu} \right) - 4\frac{1}{\cos \mu} \right\} F^{2}(1-G)^{2}\Bigg],     
 \end{split}
\end{equation}
where the pre-factor 
$\cos^{2}\mu \left(1+\frac{\sin^{2} \tilde{\theta}}{\sin^{2} \theta}\right)-2\cos \mu$ 
in the second line is clearly negative-definite for $\mu \neq 0$.
Also, the pre-factor 
$1 + \cos^{2}\mu + \frac{\sin^{2} \tilde{\theta}}{\sin^{2} \theta}\left(1+ \frac{1}{\cos^{2}\mu}\right) - 4\frac{1}{\cos \mu}$ 
in the third line is likely to be negative-definite for $\mu \neq 0$, as it vanishes for $\mu = 0$ and is approximated to be $- 4\frac{1}{\cos \mu} < 0$ for $\mu \to \pm \frac{\pi}{2}$.

\begin{figure}[h]
 \centering
 \includegraphics{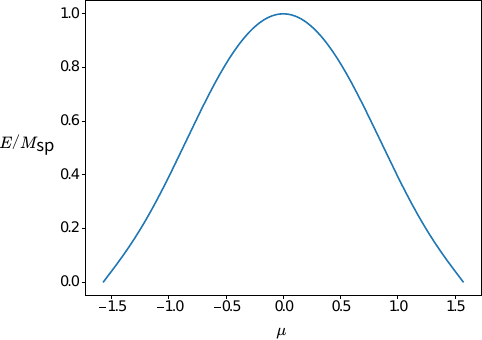}
 \label{fig-energy}
 \caption{The behaviour of $\frac{E}{M_\SP}$ as a function of the parameter $\mu$.}
\end{figure}

This expectation has been confirmed by numerical calculation. In Fig.\ref{fig-energy}, we present the result of the numerical calculation of $\frac{E}{M_\SP}$ as a function of the parameter $\mu$ for the range $- \frac{\pi}{2} \le \mu \leq \frac{\pi}{2}$.  
Now it is apparent that the sphaleron-like solution we found, \EQREF{3.17}, corresponding to $\mu = 0$,  is the maximal point of the energy under the $\mu$-dependent deformation of the fields, while $E$ vanishes for 
$\mu = \pm \frac{\pi}{2}$, as we anticipated. 

Concerning the CS number $N_\CS $, for $\mu = 0$, $N_\CS  = \frac{1}{2}$, as $\tilde{A}'_{i,y} = A^\SP _{i,y}$ in this case.
For $\mu = \pm \frac{\pi}{2}$, 
\be 
\label{4.14}  
\tilde{A}' _{y}= \tilde{A}_{y} 
= \left(-i\frac{gv}{2}\right) \sigma_{3}, \ \ \tilde{A}'_{i} = 0, 
\ee     
since $\tilde{A}_{i} = 0$ and $\tilde{V} = e^{-\pi R \tilde{A}_{y}}$ is a constant matrix in this case, leading to $N_\CS  = 0$. 
Hence, we conclude that the sphaleron-like solution we found may be regarded as a saddle point of energy, in which the energy decreases along the direction where $N_\CS $ varies.

\section{Summary and discussion}

In this paper, invoking the framework of higher dimensional gauge theory, especially the scenario of gauge-Higgs unification (GHU), we have analytically constructed a sphaleron-like solution, 
i.e. a static finite energy solution to the equation of motion in the system of SU(2) gauge field and adjoint scalar field (from the viewpoint of 4D space-time), 
whose gauge field carries the Chern-Simons number $N_\CS  = \frac{1}{2}$. 
The solution is based on the BPS monopole solution, which in turn can be understood as a self-dual field in the 4D space (not space-time) in the framework of 5D SU(2) GHU, 
once the extra space component of the gauge field $A_{y}$ is identified with the adjoint scalar needed for the construction of the 't Hooft Polyakov monopole \cite{'t Hooft_monopole, Polyakov}, 
and if we focus only on the sector of Kaluza-Klein (KK) zero mode.

By virtue of such specific property as a self-dual field, which arises only when the BPS monopole solution is embedded into the scenario of GHU, 
we achieve the construction of gauge field configuration $A_{i} \ (i = 1,2,3)$ with $N_\CS  = \frac{1}{2}$. 
This is realized by making $y$-dependent local gauge transformation in order to eliminate $A_y$ and then taking the slice at $y = 0$ of $A_{i}$, 
although we finally recover the adjoint scalar $A_y$ by switching the gauge transformation to the $y$-independent one.       
As a characteristic feature of this construction invoking higher dimensional gauge theory, in clear contrast to the case of the ordinary BPS monopole in the 3D space, 
the VEV of the adjoint scalar field is not an arbitrary parameter, but is topologically fixed, just as the action of the self-dual gauge field in the 4D space-time, such as instanton, is quantized.
Thus, the mass of the sphaleron-like solution is fixed to be 
$M_\SP  = \frac{4\pi}{g_{4}^{2}}\frac{1}{R}$ ($g_{4}$: 4D gauge coupling constant, $R$: 
the radius of the circle as the extra dimension). 
Therefore the transition probability between the vacua with different $N_\CS $ at finite temperature is predictable without uncertainty.  

We also argued that the sphaleron-like solution we found may be regarded as a saddle point of the energy in the space of the static field configurations. 
Namely, by adopting two concrete examples of the parameter-dependent field deformation, 
we have shown that the energy decreases along the direction, where the Chern-Simons number $N_\CS $ varies, just as in the case of ordinary instanton solution.   

In order to apply the method of the analytic construction proposed in this paper for the system of the SU(2) gauge field and the Higgs doublet, 
in which the original sphaleron solution was found \cite{Manton}, there remains an important issue to be settled.
As far as the GHU scenario is adopted, the scalar field inevitably belongs to the adjoint representation of the gauge group, while the Higgs doublet of course belongs to the fundamental representation.
Such difficulty can be evaded once the gauge group is enlarged and the fundamental representation of the original gauge group is identified with the \lq\lq off-diagonal\rq\rq generators of the enlarged group.
In fact, the minimal electro-weak unified SU(3) GHU model has been constructed \cite{Kubo}, where the SU(3) symmetry is broken into the gauge symmetry of the standard model, SU(2)$\times$U(1), by adopting an orbifold S$^{1}$/Z$_{2}$ as the extra space.
The KK zero modes of the components of $A_y$, belonging to the broken generators, just behaves as the Higgs doublet of the standard model.
So it is an interesting possibility that the original sphaleron solution can be reconstructed analytically in the framework of the SU(3) GHU with orbifold extra space, though its investigation is beyond the scope of the present paper and we would like to leave it for future research.

\subsection*{Acknowledgments}

This work was supported in part by Japan Society for the Promotion of Science, Grants-in-Aid for Scientific Research, No.~15K05062.

%\bibliography{./GHU_ref_library} 
%\bibliographystyle{./JHEP}

%%%%%%%%%%%%%%%%%%%%
%%%% From here
%%%%%%%%%%%%%%%%%%%%
\providecommand{\href}[2]{#2}\begingroup\raggedright\endgroup
%%%%%%%%%%%%%%%%%%%%
%%%% Till here
%%%%%%%%%%%%%%%%%%%%

\end{document}